\documentclass[conference,10pt]{IEEEtran}
\PassOptionsToPackage{bookmarks={false}}{hyperref}
\PassOptionsToPackage{draft}{hyperref}
\usepackage{tikz}
\usepackage{pgfplots}
\usepackage{amsmath,epsfig,amssymb,algorithm,algpseudocode,amsthm,cite,url}
\usepackage{hyperref}
\usepackage{enumerate}
\usepackage[latin1]{inputenc}

\DeclareMathOperator{\diag}{diag}
   
\theoremstyle{plain}

\theoremstyle{definition}

\theoremstyle{remark}
\newtheorem{rem}{Remark}

\newtheoremstyle{specialcasestyle}{1mm}{1mm}{\upshape}{}{\bfseries\upshape}{.}{0mm}{}
\theoremstyle{specialcasestyle}

     \newcommand{\br}{{\bf r}}
      \newcommand{\bU}{{\bf U}}

  \newcommand{\bA}{{\bf A}}
  \newcommand{\bB}{{\bf B}}

    \newcommand{\bI}{{\bf I}}
      \newcommand{\bh}{{\bf h}}
      \newcommand{\bH}{{\bf H}}

         \newcommand{\ba}{{\bf a}}

      \newcommand{\bs}{{\bf s}}
       
          \newcommand{\be}{{\bf e}}
          \newcommand{\bz}{{\bf z}}
      
        \newcommand{\bp}{{\bf p}}
        
      \newcommand{\bw}{{\bf w}}
      
        \newcommand{\bW}{{\bf W}}

\usepackage{graphicx}
\usepackage{color}
\usepackage{cite}
\usepackage{placeins}
\usepackage{float}
\usepackage{tabularx,colortbl}
\usepackage[font=normalsize]{caption}
\IEEEoverridecommandlockouts
\begin{document}
\title{
Optimal Linear Precoding for Indoor Visible Light Communication System}
\author{\IEEEauthorblockN{Houssem Sifaou, Ki-Hong Park, Abla Kammoun, Mohamed-Slim Alouini}

\IEEEauthorblockA{
 Computer, Electrical, and Mathematical Sciences \& Engineering Division, \\King Abdullah University of Science and Technology, Thuwal, Saudi Arabia\\}
\vspace{-0.8cm}}

\maketitle

\begin{abstract}

Visible light communication (VLC) is an emerging technique that uses light-emitting diodes (LED) to combine communication and illumination. It is considered as a promising scheme for indoor wireless communication that can be deployed at  reduced costs while offering high data rate performance. In this paper, we focus on the design of the downlink of a multi-user VLC system. Inherent to multi-user systems is the interference caused by the broadcast nature of the medium. Linear precoding based schemes are among the most popular solutions that have recently been proposed to mitigate inter-user interference. 
This paper focuses on the design of the optimal linear precoding scheme that solves the max-min signal-to-interference-plus-noise ratio (SINR) problem. The performance of the proposed precoding scheme is studied under different working conditions and compared with the classical zero-forcing precoding. Simulations have been provided to illustrate the high gain of the proposed scheme. 
\end{abstract}


\IEEEpeerreviewmaketitle

\section{Introduction}

Recently, we have witnessed an increasing interest in the technology of visible light communications (VLC) \cite{Tsonev2014,Tsonev2015}, as a result of the recent advances in the fabrication of light emitting diodes (LEDs). VLC uses white LEDs that transmit data by changing the light intensity; but variations of the modulated optical signal  cannot be noticed by the human eyes which perceive only the average light intensity. Due to its main advantages, such as ease of deployment and
low cost, VLC is now being considered  as a potential candidate to complement conventional indoor radio frequency (RF) communications.

The use of multiple-input-multiple-output (MIMO) techniques  appears to be natural in VLC systems as illumination usually requires the use of multiple LEDs.  This excess in the number of degrees of freedom offered by the availability of multiple transceivers can be leveraged to ensure high data rates \cite{Takase2004,Zeng2009a,Yu2013,Li2015a}. Very recently, the use of multi-user MIMO (MU-MIMO) techniques for VLC systems has been studied \cite{Yu2013,Li2015a}, where the issue of mitigating the inter-user interference has
been addressed. Towards this goal, linear precoding schemes aiming at minimizing  the mean square error (MSE) \cite{Li2015a} or maximizing the signal-to-interference-plus-noise ratio (SINR) \cite{Yu2013} have been proposed. 

In this paper, we consider the problem of designing the optimal linear precoding (OLP) that solves the max-min SINR problem in  MU-MIMO VLC systems. Such a problem has been widely investigated in RF MU-MIMO systems \cite{Cai2012,Sifaou2016a,Sifaou2016b,Cumanan2010}. However, the results of these works could not be applied since they do not take into consideration many practical considerations of VLC systems. As a matter of fact,  time-domain signals in VLC are real-valued and positive, while RF counterparts are complex. Moreover, a constraint on the average optical power should be considered in VLC to meet with technical illumination requirements.    

This paper derives the optimal linear precoding that solves the max-min SINR problem and  compare its performance with the previously proposed Zero forcing (ZF) precoding in  \cite{Yu2013}. We show that our proposed precoding provides a significant gain in performance especially when the users are close to each other and  the inter-user interference is high.

The remainder of the paper is organized as follows. The next section introduces the system model. In section \ref{Precoding_Design}, the OLP is designed. Before concluding in section \ref{conclusion}, numerical results are given in section \ref{Numerical_Results}.

\begin{figure}[]
 \label{figg11}
  \centering
    \includegraphics[scale=0.45]{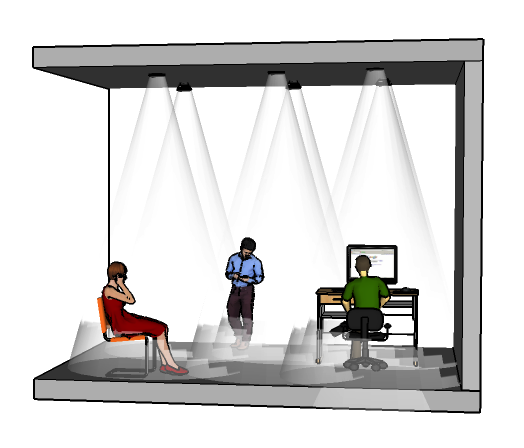}
    \vspace{-0.5cm}
     \caption{Indoor VLC system}
\end{figure}


\section{System Model}
\label{system_model}
We consider a MU-MIMO VLC system where $M$ transmitters communicate with $K$ user equipments (UEs)$(M>K)$ as shown in Fig. 1. Each UE is equipped with single receiving unit. Intensity modulation is employed at the transmitter and direct detection at the receiver. The transmitting LEDs produce light intensity proportional to the input electric signal and the receiver converts the received light intensity into electric signal. The light intensity variation can not be detected by the human eyes which perceive only the average light intensity. Taking into account the required level of illumination, the characteristics of LEDs and the presence of intensity modulation, three constraints on the input electric signal $y_n$ have to be considered:

\begin{itemize}
\item The input electric signal must be real valued and positive: $y_n\ge0$ .
\item The input electric signal $y_n$ must be lower than a value $p_{max}$ in order to ensure that the LED works in its linear dynamic range \cite{Li2015a}:  $y_n\le p_{max}$.
\item The expectation of the input electric signal must be equal to a constant determined by the illumination level: $\mathbb{E}(y_n)=p_n,\ \ \forall n=1,\cdots,M$ \cite{Wang2013a}.
\end{itemize}

Let $s_k$ be the symbol intended to UE $k$. We assume that $s_k\in[-1,1]$ with zero mean. Let $\bW=[w_{n,k}] \in \mathbb{R}^{M\times K}$ denotes the precoding matrix. The transmitted signal at the $n$-th transmitting unit is:
\begin{equation}
x_n=\sum_{k=1}^{K}s_k w_{n,k}.
\end{equation}
In order to satisfy the constraint $\mathbb{E}(y_n)=p_n,\ \ \forall n$, a DC offset $p_n$ should be added to $x_n$.
\begin{equation}
y_n=x_n+p_n.
\end{equation}
Since $s_k\in[-1,1]$, we have
\begin{equation}
-\sum_{k=1}^{K}|w_{n,k}|+p_n\leq y_n \leq \sum_{k=1}^{K}|w_{n,k}|+p_n, \ \ \forall n.
\end{equation}
To ensure that $y_n$ is in the linear dynamic range of the LEDs, we should satisfy 
\begin{equation}
 \sum_{k=1}^{K}|w_{n,k}|+p_n\leq p_{max}.
 \label{dynamic_range}
 \end{equation}
Under the assumption of positivity of the input signal 
\begin{equation}
 -\sum_{k=1}^{K}|w_{n,k}|+p_n\geq 0.
 \label{positivity}
 \end{equation}
The two constraints in \eqref{dynamic_range} and \eqref{positivity} can be combined as:
\begin{equation}
\sum_{k=1}^{K}|w_{n,k}|\leq\tilde p_n,
\label{power_constraint}
 \end{equation}
where $\tilde p_n={\rm min}(p_n,p_{max}-p_n)$.

For VLC, the line of sight propagation dominates \cite{Zeng2009a,Komine2004a} and the channel gain from the $n$-th transmitter to the $k$-th receiver can be expressed as \cite{Zeng2009a,Khan1997}:
\begin{align}
h_{n,k}=
\left\lbrace
\begin{array}{r c l}
 \frac{\rho A}{i_{n,k}^2}R(\phi_{n,k})\cos(\theta_{n,k}), & \  \  &\theta_{n,k}\leq \theta_{c} \\
0, & \  \  &\theta_{n,k}> \theta_{c} 
\end{array}
\right.
\end{align}
where $\phi_{n,k}$ is the angle of emission with respect to the $n$-th transmitter, $\theta_{n,k}$ is the incident angle with respect to the $k$-th receiver, $i_{n,k}$ is the distance that separates the $n$-th transmitter and the $k$-th receiver, $\theta_{c}$ is the receiver filed of view (FOV), $\rho$ is the photo detector (PD) responsivity and $A$ is the collection area given by:
$$
A=\frac{q^2}{\sin^2(\theta_c)}A_{PD}
$$
where $q$ is the refractive index of optical concentrator and $A_{PD}$ is the PD area. $R(\phi_{n,k})$ is the Lambertian radiant intensity:
$$
R(\phi_{n,k})=\frac{(m+1)\cos^m(\phi_{n,k})}{2 \pi},
$$
where $m$ is the order of the Lambertian emission mode number \cite{Zeng2009a,Khan1997}.
Let $\bh_k=[h_{1,k},\cdots,h_{M,k}]^T$ be the channel vector corresponding to the $k$-th UE. After removing the DC offset $p_n$ introduced at the transmitter, the useful received signal at the $k$-th UE can be expressed as:
\begin{equation}
r_k=\bh_k^T\bw_ks_k+\sum_{j\neq k}\bh_k^T\bw_j s_j+z_k,
 \end{equation}
where $z_k$ is the additive noise. In VLC, $z_k$ is assumed to be real valued Gaussian distributed with zero mean and variance $\sigma_k^2$ \cite{Zeng2009a,Khan1997}:
$$
\sigma_k^2=2eP_{s,k} B+ 2e \rho \xi_{amb} A 2 \pi (1-\cos(\theta_c))B +i_{amp}^2B
$$
where $e$ is the electronic charge, $B$ is the bandwidth, $\xi_{amb}$ is the ambient photocurrent, $i_{amp}$ is the preamplifier noise density and $P_{s,k}$ is defined as:
$$
P_{s,k}=\sum_{n=1}^{M}p_n h_{n,k}.
$$
For notational convenience, we define $\boldsymbol{\sigma}=[\sigma_1,\cdots,\sigma_K]^T$.
The vector $\br$ collecting the received signals at the UEs can be expressed as:
$$
\br=\bH\bW\bs+\bz
$$
where $\bH \in \mathbb{R}^{K\times M}$ is the channel matrix, $\bs=[s_1,\cdots, s_K]^T$ denotes the symbol vector and $\bz=[z_1,\cdots, z_K]^T$ is the noise vector.


\section{Precoding Design}
\label{Precoding_Design}
In this section, we address the problem of designing the optimal linear precoding (OLP) that solves the max-min SINR problem while satisfying the optical power constraint in \eqref{power_constraint}. Due to complexity concerns, the linear precoding scheme are generally preferred to the nonlinear ones. 
Prior to presenting our proposed scheme, we shall review the classical zero-forcing (ZF) scheme, which will be used later for comparison. 


\subsection{Zero-forcing (ZF):}
The ZF precoding matrix is defined as:
$$\bW=\bH^T \left( \bH \bH^T \right)^{-1} diag(\boldsymbol{\gamma}),$$ 
where $\boldsymbol{\gamma}=[\gamma_1,\cdots,\gamma_K]^T$ and $\gamma_k$ is the symbol gain for UE $k$. The ZF precoding suppresses the interference and the received signal can be expressed as:
$$
r_k=\gamma_k s_k+z_k,
$$
The optimal symbol gain vector $\boldsymbol{\gamma}$ has been determined in \cite{Yu2013} where it has been shown that:
$$
\gamma_k^\star={\sigma_k}\mu_k^\star,
$$
where $\mu_k^\star=\underset{n}\min \frac{\tilde p_n}{(\bA \boldsymbol{1}_K)n}$ and $\bA={\rm abs}(\bH^T \left( \bH \bH^T \right)^{-1} )\diag(\boldsymbol{\sigma})$. $ \boldsymbol{1}_K$ is the all-one vector of size $K$.


\subsection{Optimal linear precoding:}
In this section, we propose to determine the optimal linear precoding that solves the following max-min SINR problem:
\begin{align*}
	{\mathcal{P}}:  \quad \max_{\bW}\ \ \min_{k} &  \ \ {\rm SINR}_k\\
\textnormal{subject to} \quad&  \sum_{k=1}^{K}|w_{n,k}|\leq \tilde p_n, \ \ n=1,\cdots,M
\end{align*}
where ${\rm SINR}_k$ is given by:
$$
{\rm SINR}_k=\frac{|\bh_k^T\bw_k|^2}{\sum_{j\neq k}|\bh_k^T\bw_j|^2+\sigma_k^2}.
$$
Problem $({\mathcal{P}})$ can be rewritten as
\begin{align*}
	{\mathcal{P}}_1:  \quad \max_{\bW, t} \ \  t\\
\textnormal{subject to} \quad&  \sum_{k=1}^{K}|w_{n,k}|\leq \tilde p_n, \ \ n=1,\cdots,M\\
& \frac{|\bh_k^T\bw_k|^2}{\sum_{j\neq k}|\bh_k^T\bw_j|^2+\sigma_k^2}\geq t, \ \ k=1, \cdots,K.
\end{align*}
To solve $\mathcal{P}_1$, we shall first rewrite the constraints in a different form. 
The power constraints in $({\mathcal{P}}_1)$ can be rewritten as :
$$
{\|\bW^T \be_n \|}_{1} \leq \tilde p_n, \ \ n=1,\cdots,M,
$$
where $\be_n$ is the all zero vector having the $n$-th element equal to 1. The $L_1$ norm constraints can be transformed into the following linear constraints:
\begin{equation}
-\ba_n \leq \bW^T \be_n\leq \ba_n, \quad   \boldsymbol{1}_K^T \ba_n \leq \tilde p_n, \ \ n=1,\cdots,M,
\label{pe}
\end{equation}
where $\ba_n \in \mathbb{R}^K$ is a new optimization variable. In order to simplify further the optimization problem, we introduce the following vectors : $\bw=vec(\bW^T)=[\be_1^T\bW,\cdots,\be_M^T\bW]^T$ and $\ba=[\ba_1^T, \cdots,\ba_M^T]^T$. Using these notations, the power constraints in \eqref{pe} can be written as:
$$
-\ba\leq \bw \leq \ba, \quad \bU \ba \leq \tilde \bp,
$$
where $\bU =\bI_M \otimes \boldsymbol{1}_K^T$ and $\tilde \bp=[\tilde p_1,\cdots,\tilde p_M]^T$, $\bI_M$ being the identity matrix of size $M$ and $\otimes$ denoting  the Kronecker product.
We  now work out the SINR constraints of $({\mathcal{P}}_1)$:
$$
\sum_{j\neq k}|\bh_k^T\bw_j|^2+\sigma_k^2\leq \frac{1}{t} |\bh_k^T\bw_k|^2, \ \ \forall k,
$$
 to express them in terms of vector $\bw$. 
 The left hand side can be rewritten using $L_2$ norm as: {\footnote{Without loss of optimality, we can assume that $\bh_k^{T}\bw_k \geq 0$ since the objective function and the constraints are invariant to sign changes of ${\bf w}_k$.}}
\begin{equation}
\left\| \begin{array}{l} \bW_k^T \bh_k \\ \sigma_k \end{array}\right\|_2 \leq \frac{1}{\sqrt{t}} \bh_k^T\bw_k,\ \ \forall k, 
\label{SINR_cons}
\end{equation}
where $\bW_k$ is the matrix obtained from $\bW$ by removing the $k$-th column. Let $\bI_K^k$ denotes the matrix obtained from identity matrix of size $K$ by setting the $(k,k)$-th element to zero. Then:
\begin{align*}
    \bW_k^T \bh_k&=\bI_K^k \bW^T\bh_k\\
                  &={\rm vec}(\bI_K^k \bW^T\bh_k) \\
                  &=(\bh_k^T \otimes \bI_K^k) \bw.
\end{align*}
and
\begin{align*}
    \bw_k&=vec(\bw_k^T)=vec(\be_k^T\bW)\\
         &=(\bI_M\otimes \be_k^T) vec( \bW^T)=(\bI_M\otimes \be_k^T) \bw.
\end{align*}
 Thus, the constraints in \eqref{SINR_cons} can be reformulated as:
\begin{equation}
\|   \bB_k \bw + \boldsymbol{\sigma}_k   \|_2\leq \frac{1}{\sqrt{t}}  \bh_k^T(\bI_M\otimes \be_k^T) \bw, \ \ \forall k,
\label{second_order_cone_cons}
\end{equation}
with $\bB_k \in \mathbb{R}^{(K+1)\times MK}$ and $\boldsymbol{\sigma}_k \in \mathbb{R}^{(K+1)} $ are given by:
\begin{align*}
\bB_k =\left[\begin{array}{l} \bh_k^T \otimes \bI_K^k  \\ \boldsymbol{0}_{1\times MK}  \end{array} \right], \quad \boldsymbol{\sigma}_k =[0, \cdots, 0, \sigma_k]^T.
\end{align*}
For  fixed $t$, the reformulated SINR constraints in \eqref{second_order_cone_cons} are second-order cone constraints which are convex \cite{Wiesel2006,bengtsson01,YuL07a}. Our optimization problem ${\mathcal{P}}_1$ turns out to be quasi-convex and can be solved using the bisection algorithm \cite{boyd}. Each iteration of the bisection algorithm consists in holding $t$ fixed and solving for ${\bf w}$ the following feasibility problem,
\begin{equation}
\begin{aligned}
	 {\rm find} \ \ \bw\\
\textnormal{subject to} \ \ &  \|   \bB_k \bw + \boldsymbol{\sigma}_k   \|_2\leq \frac{1}{\sqrt{t}}  \bh_k^T(\bI_M\otimes \be_k^T) \bw, \ \ \forall k\\
&-\ba\leq \bw \leq \ba, \quad \bU \ba \leq \tilde \bp.
\end{aligned}
\label{feas_prob}
\end{equation}
which is a  second-order cone  program (SOCP) that can be efficiently solved using CVX \cite{cvx}. 
The optimal $t$ corresponds thus to the maximum value
for which it exists ${\bf w}$ satisfying the constraints in \eqref{feas_prob}. To sum up, solving $\mathcal{P}_1$ can be performed using the following algorithm:  \begin{algorithm}[H]
\label{alg}
\caption{Iterative algorithm for computation of OLP}
\begin{algorithmic}
\State Let $t_1>0$ and $t_2>0$ such that $t_1<t_2$ and problem \eqref{feas_prob} is infeasible when $t=t_2$ and feasible when $t=t_1$.
\State Initialize the precision parameter $\epsilon$.
\While{ $t_2-t_1>\epsilon$}{
\State $t=(t_1+t_2)/2$
\State Solve the feasibility problem \eqref{feas_prob}.
\If {the problem is feasible} $t_1=t$ \Else \ \ $t_2=t$ \EndIf}
\EndWhile
\State The optimal precoding vector of the $k$-th UE is $\bw_k^\star=(\bI_M\otimes \be_k^T) \tilde \bw$ where $\tilde \bw$ is the last feasible solution to \eqref{feas_prob}.
\end{algorithmic}
\end{algorithm}
Algorithm 1 converges in exactly $\log_2[(t_2-t_1)/\epsilon]$ iterations. In order to accelerate the convergence of Algorithm 1, the initial values $t_1$ and $t_2$ can be determined efficiently using the following algorithm:
\begin{algorithm}[H]
\label{alg_2}
\caption{Iterative algorithm for the computation of the initial values $t_1$ and $t_2$.}
\begin{algorithmic}
\State Set $t_{lower}=10^{-5}$ (we know that the optimal $t$ is positive and we choose very small value to ensure feasibility).
\Repeat
\State $t_{upper}=\alpha t_{lower}$.
\State Solve the feasibility problem \eqref{feas_prob} with $t=t_{upper}$.
\If {the problem is feasible} $t_{lower}=t_{upper}$ \EndIf
\Until{$t_{lower}\neq t_{upper}$}
\State  Set $t_1=t_{lower}$ and $t_2=t_{upper}$.
\end{algorithmic}
\end{algorithm}
where $\alpha$ is a scale factor strictly greater than 1. 
\begin{rem}
    The overall complexity of the precoding design is governed by the complexity of Algorithm 1 and Algorithm~2. For Algorithm 1, the number of iterations is given by  $\log_2[(t_2-t_1)/\epsilon]$ and as such is low when the difference $t_2-t_1 $ is small. On the other hand, Algorithm 2 would require less iterations as  $\alpha$ increases. However, increasing $\alpha$ produces higher values for the difference  $t_2-t_1 $, and as a consequence increases the complexity of Algorithm 1.
    That being said, it is worth mentioning that    Algorithm 1 and Algorithm 2 are used at the pace of the change of the channel, which is slow as far as VLC applications are concerned.  
    \end{rem}

\section{Numerical Results}
\label{Numerical_Results}
\begin{figure}[]
\begin{center}
    \includegraphics[width=0.5\textwidth]{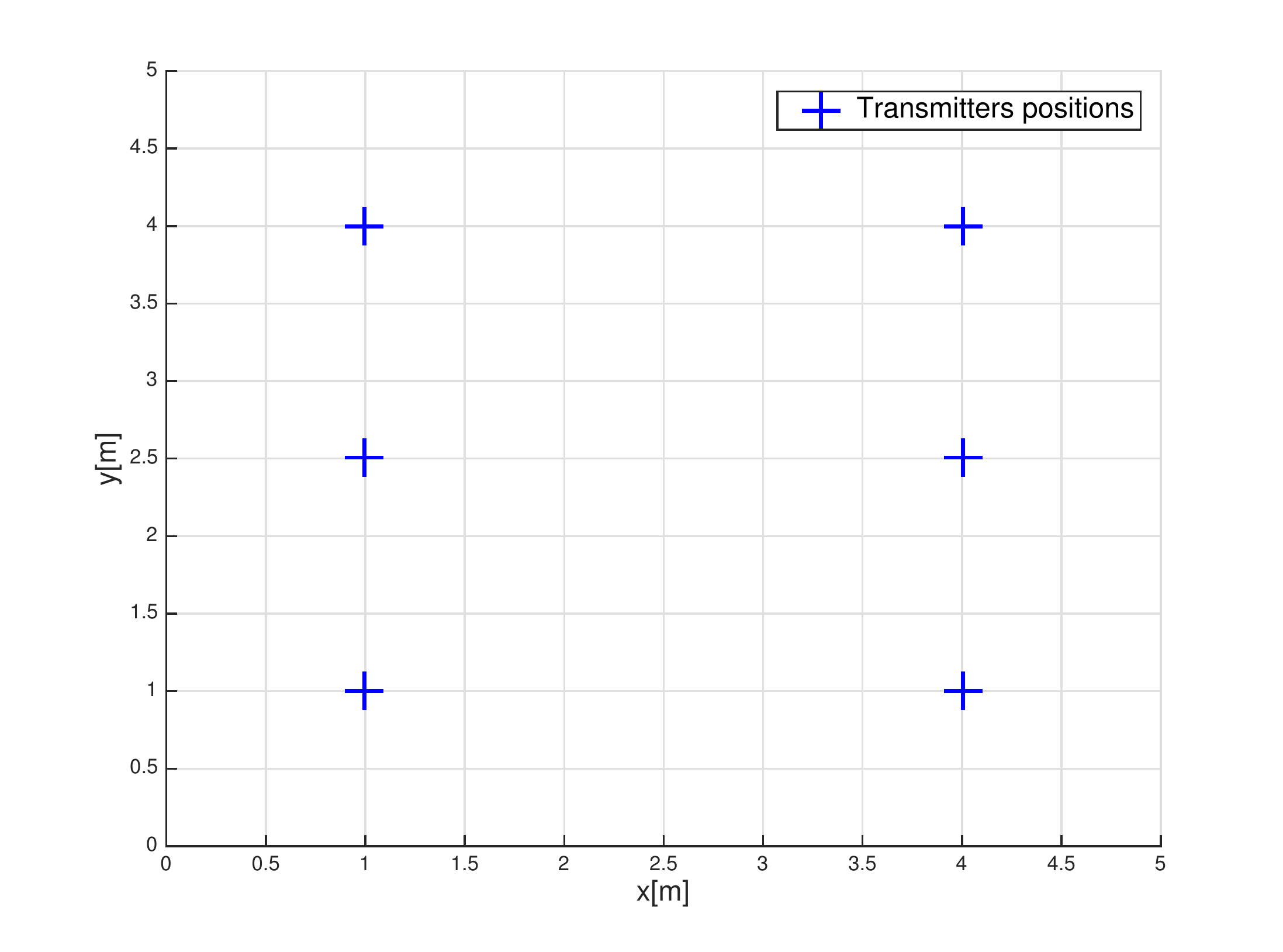}
    \vspace{-0.8cm}
     \label{pos_LED_array}
     \caption{Positions of the transmitters on the roof.}
     \end{center}
\end{figure}

In this section, the proposed OLP is compared with the classical ZF precoding using simulations. We consider a  VLC system  composed of 6 transmitters installed on the ceiling of the room as depicted  in Fig. 1. The positions of the transmitters are given in Fig. 2. We assume that  UEs are static, which allow us to consider static channels.  The average powers for all transmitters  $\left\{p_n\right\}$ are assumed to be the same and equal to $p$ where, for simulation purposes, $p$ is assumed to range between $15{\rm dBm}$ and $30{\rm dBm}$. Without loss of
generality, we assume also that $p_{max}-p_n \gg p_n$ and as such $\tilde p_n=\min(p_n,p_{max}-p_n )=p_n$. The VLC system parameters are summarized in Table 1. The performance measure is the rate per UE $r$ defined as:
$$
r=\frac{1}{K}\sum_{k=1}^{K}B\log_2\left(1+{\rm SINR}_k\right).
$$

 \begin{table}[H]
    \caption{VLC system parameters.}
    \begin{center}
    \begin{tabular}{| l | l |}
    \hline
    Room Size & $5m \times 5m \times 3m$  \\ \hline
    Mode number $m$ of Lambertian emission & 1 \\ \hline
    Photo Detector reponsivity $\rho$& 0.4 A/W \\\hline
    Photo Detector area $A_{PD}$ & 1 $cm^2$ \\\hline
    Receiver FOV $\theta_c$ &60 deg\\\hline
    Refractive index of optical concentrator $q$ &1.5 \\\hline
    Pre-amplifier noise density $i_{amp}$ & 5 $pA/Hz^{-1/2}$ \\\hline
    Ambient light photocurrent $\xi_{amb}$ & 10.93 $A/m^2/Sr$ \\\hline
    Bandwidth $B$ & 100$MHz$ \\\hline
    \end{tabular}
    \label{table11}
    \end{center}
    \end{table}
 
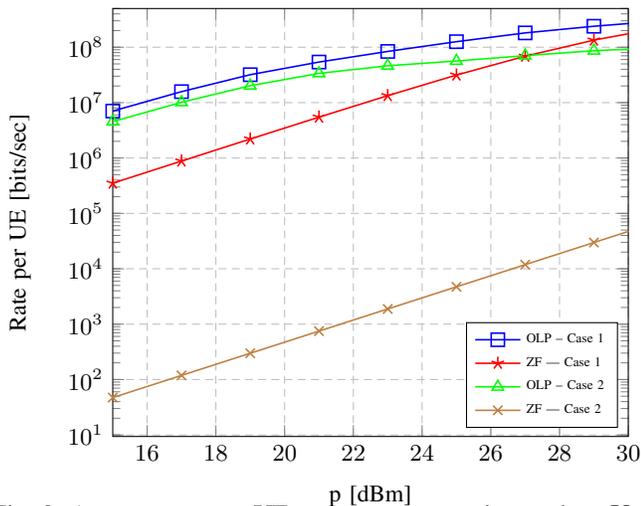
\begin{figure}
  \centering
   \begin{tikzpicture}[scale=1,font=\small]
    \renewcommand{\axisdefaulttryminticks}{4}
    \tikzstyle{every major grid}+=[style=densely dashed]
    \tikzstyle{every axis y label}+=[yshift=-10pt]
    \tikzstyle{every axis x label}+=[yshift=5pt]
    \tikzstyle{every axis legend}+=[cells={anchor=west},fill=white,
        at={(0.98,0.02)}, anchor=south east, font=\tiny ]
    \begin{semilogyaxis}[
      xmin=15,
      ymin=0,
      xmax=30,
      ymax=5*10^8,
      grid=major,
      scaled ticks=true,
   			xlabel={p [dBm]},
   			ylabel={Rate per UE [bits/sec]}			]
   \addplot[color=blue,mark size=2.5pt,mark=square,line width=0.6pt] coordinates{    
(15,7.040075e+06)(17,1.577842e+07)(19,3.193846e+07)(21,5.400667e+07)(23,8.394340e+07)(25,1.260024e+08)(27,1.814833e+08)(29,2.393200e+08)(31,3.000868e+08)(33,3.806610e+08)(35,4.862679e+08)
         };                               \addlegendentry{OLP -- Case 1}
                                         
                  \addplot[color=red,mark size=2.5pt,mark=star,line width=0.6pt] coordinates{    
(15,3.512516e+05)(17,8.806742e+05)(19,2.201984e+06)(21,5.468418e+06)(23,1.336076e+07)(25,3.146586e+07)(27,6.889512e+07)(29,1.343259e+08)(31,2.280992e+08)(33,3.418352e+08)(35,4.662400e+08)         };                              \addlegendentry{ZF --- Case 1} 
  \addplot[color=green,mark size=2.5pt,mark=triangle,line width=0.6pt] coordinates{    
(15,4.545353e+06)(17,1.014447e+07)(19,2.019833e+07)(21,3.370525e+07)(23,4.622874e+07)(25,5.644082e+07)(27,7.048550e+07)(29,8.651884e+07)(31,9.853847e+07)(33,1.054377e+08)(35,1.131936e+08)         };                              \addlegendentry{OLP -- Case 2}
                                         
                  \addplot[color=brown,mark size=2.5pt,mark=x,line width=0.6pt] coordinates{    
(15,4.724389e+01)(17,1.186698e+02)(19,2.980788e+02)(21,7.487153e+02)(23,1.880586e+03)(25,4.723396e+03)(27,1.186284e+04)(29,2.979031e+04)(31,7.479453e+04)(33,1.877077e+05)(35,4.706640e+05)         };
                              \addlegendentry{ZF --- Case 2}                                                                 
                                    \end{semilogyaxis}
  \end{tikzpicture} \vskip-4mm
\centering
  \caption{Average rate per UE vs. power constraint $p$ when $K=4$.}
  \label{fig2}
\end{figure}

    In Fig. \ref{fig2}, we compare the rate per UE vs. the power constraint of the proposed OLP with that of the classical ZF precoding for two different scenarios depending on the separation between the UEs. Namely, the first scenario corresponds to the case where the UEs are positioned far away from each other, while the second scenario studies the situation in which UEs become closer. The positions of UEs in both scenarios are indicated in Table 2 and Table 3 respectively.  As seen
    from Fig. \ref{fig2}, the performance of the OLP is better and the improvement in performance becomes higher when the UEs are close to each other or the power constraint $p$ is low (low SNR regime).

\begin{figure}
  \centering
   \begin{tikzpicture}[scale=1,font=\small]
    \renewcommand{\axisdefaulttryminticks}{4}
    \tikzstyle{every major grid}+=[style=densely dashed]
    \tikzstyle{every axis y label}+=[yshift=-10pt]
    \tikzstyle{every axis x label}+=[yshift=5pt]
    \tikzstyle{every axis legend}+=[cells={anchor=west},fill=white,
         at={(0.98,0.02)}, anchor=south east, font=\tiny ]
    \begin{semilogyaxis}[
      xmin=15,
      ymin=0,
      xmax=30,
      ymax=5*10^8,
      grid=major,
      scaled ticks=true,
   			xlabel={p [dBm]},
   			ylabel={Average per UE rate [bits/sec]}			]
   \addplot[color=blue,mark size=2pt,mark=square,line width=0.6pt] coordinates{    
(15,1.277580e+07)(17,2.793102e+07)(19,5.244302e+07)(21,8.304762e+07)(23,1.199365e+08)(25,1.688881e+08)(27,2.325877e+08)(29,3.179037e+08)(31,4.220408e+08)(33,5.407978e+08)(35,6.665347e+08)

         };
                              \addlegendentry{ OLP ---3 UEs}
                                         
                  \addplot[color=red,mark size=2pt,mark=star,line width=0.6pt] coordinates{    
(15,9.001799e+05)(17,2.250542e+06)(19,5.587686e+06)(21,1.364444e+07)(23,3.209435e+07)(25,7.011278e+07)(27,1.362700e+08)(29,2.306545e+08)(31,3.447700e+08)(33,4.693847e+08)(35,5.987439e+08)

         };
                              \addlegendentry{ ZF ---3 UEs}  
                              \addplot[color=green,mark size=2pt,mark=triangle,line width=0.6pt] coordinates{    

(15,6.594084e+06)(17,1.448174e+07)(19,2.728808e+07)(21,4.463535e+07)(23,6.701397e+07)(25,8.995049e+07)(27,1.122572e+08)(29,1.363739e+08)(31,1.769579e+08)(33,2.355255e+08)(35,3.126706e+08)

         };
                              \addlegendentry{ OLP ---5 UEs}       
                              
                        \addplot[color=brown,mark size=2pt,mark=x,line width=0.6pt] coordinates{                                        
                             (15,6.679436e+04)(17,1.677184e+05)(19,4.209089e+05)(21,1.054909e+06)(23,2.635134e+06)(25,6.529277e+06)(27,1.586914e+07)(29,3.695959e+07)(31,7.933908e+07)(33,1.506074e+08)(35,2.490046e+08)

        };
                              \addlegendentry{ ZF ---5 UEs}         
                                                           
                                    \end{semilogyaxis}
  \end{tikzpicture} \vskip-4mm
\centering
  \caption{Average rate per UE vs. power constraint $p$ when $K=3$ and $K=5$.}
  \label{fig3}
\end{figure}
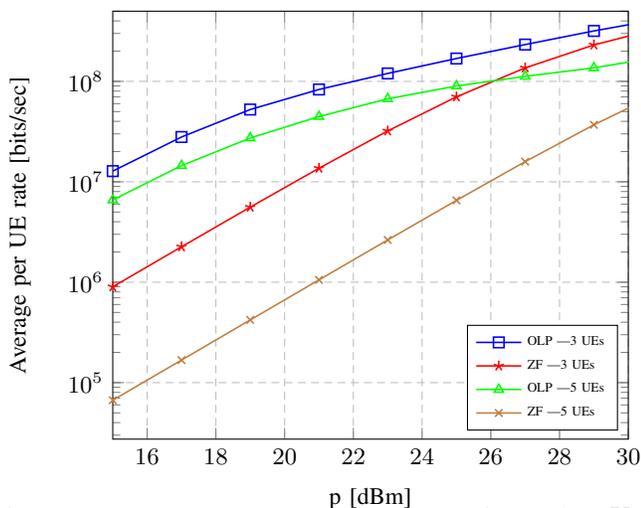

    The impact of the number of UEs is investigated in Fig. \ref{fig3}, where the rate per UE vs. $p$ is plotted for up to 5 UEs positioned randomly in the room. As seen, regardless of the number of UEs, our proposed precoding achieves a significant gain as compared to the ZF precoding especially in the low SNR regime.  
 
\begin{table}[H]
    \caption{UEs with large separation -- Case 1.}
    \begin{center}
    \begin{tabular}{| l | l |}
    \hline
    UE number & $[X \ \  \ \ \ \ Y\ \  \ \  Z]$  \\ \hline
    UE 1 & $[2.05 \ \  1.60 \ \  2.15]$  \\ \hline
    UE 2 & $[2.15  \ \ 4.10 \ \ 2.15]$  \\ \hline
    UE 3 & $[3.50 \ \ 3.50 \ \ 2.50]$  \\ \hline
    UE 4 & $[4.20 \ \  4.20 \ \  2.50]$  \\ \hline
    \end{tabular}
    \label{table11}
    \vspace{-0.5cm}
    \end{center}
    \end{table}
\begin{table}[H]
    \caption{UEs with small separation -- Case 2.}
    \begin{center}
    \begin{tabular}{| l | l |}
    \hline
    UE number & $[X \ \  \ \ \ \ Y\ \  \ \  Z]$  \\ \hline
    UE 1 & $[2.05 \ \ 2.20 \ \  2.15]$  \\ \hline
    UE 2 & $[2.05 \ \ 2.40 \ \  2.15]$  \\ \hline
    UE 3 & $[2.05 \ \ 2.60 \ \  2.15]$  \\ \hline
    UE 4 & $[2.05 \ \ 2.80 \ \  2.15]$  \\ \hline
    \end{tabular}
    \label{table11}
    \end{center}
    \end{table}


\section{Conclusion}
\label{conclusion}
In this paper, we considered the problem of precoding design for indoor VLC systems. We have determined the optimal linear precoding that solves the max-min SINR problem. It has been shown by simulation that our proposed technique provides a significant gain in performance compared to the classical ZF precoding especially in the scenario where the UEs are close to each other.

\bibliographystyle{IEEEbib}
\bibliography{IEEEabrv,IEEEconf,tutorial_RMT}
\end{document}